\documentstyle[preprint,aps,floats]{revtex}
\input psfig.tex
\begin{document}

\draft

\title{In Support of Inflation\footnote{{\sc Perspective: Cosmology},
                               published in Science 291, 837-838 (2001) 
                               (2 February 2001 issue). }}
% 
% 
%\title{New Cosmic Microwave Background Results Back Up Inflationary Paradigm}
\author{Alejandro Gangui\footnote{The author is at {\sc IAFE}, the Institute
                         for Astronomy and Space Physics, Ciudad Universitaria, 
                         1428 Buenos Aires, Argentina.  
                         E-mail: gangui@iafe.uba.ar }}

\maketitle
\hspace{0.2in}

What created the initial inhomogeneities in the Universe that resulted
in galaxies, clusters of galaxies and other large-scale structures?
This problem continues to puzzle cosmologists. But whatever the
mechanism, it must have left its signature in the cosmic microwave
background (CMB) radiation\cite{Scott_etal_Science}.  The CMB is a
relic of the big bang, a cold bath of light just a few degrees above
absolute zero
that pervades the entire the Universe.  Released when matter began to
become structured, the CMB is our earliest ``snapshot'' of the
Universe. Variations (or anisotropies) in its effective temperature
tell us about the size and strength of the initial seeds in the
primordial plasma, those clouds of gas that clumped together under
gravitational attraction and led to the birth of galaxies.  Recent CMB
experiments suggest that these fundamental seeds could have resulted
from tiny primordial quantum fluctuations generated in the early
Universe during a period of rapid (faster than light) expansion called
inflation.

Early on, when the Universe was small and very hot, the free electron
density was so high that photons could not propagate freely without
being scattered by electrons. Ionized matter, electrons and radiation
formed a single fluid, with the inertia provided by the baryons and
the radiation pressure given by the photons.  And this fluid supported
sound waves. In fact, the gravitational clumping of the effective mass
in the perturbations was resisted by the restoring radiation pressure,
resulting in gravity-driven {\it acoustic oscillations} in both fluid
density and local velocity.

As the Universe expanded and ambient temperatures decreased,
high-energy collisions became less and less frequent.  Very energetic
photons were not statistically significant to destroy the increasing
number of neutral particles (mostly hydrogen) that began to
combine. Cosmologist refer to this period as {\it recombination}. Soon
afterwards the CMB was released free, making its {\em last scattering}
upon matter.  This is a remarkable event in the history of the
Universe, because it is the very moment when it passed from being
opaque to being transparent to electromagnetic radiation.

\begin{figure}[t]
\centerline{\psfig{file=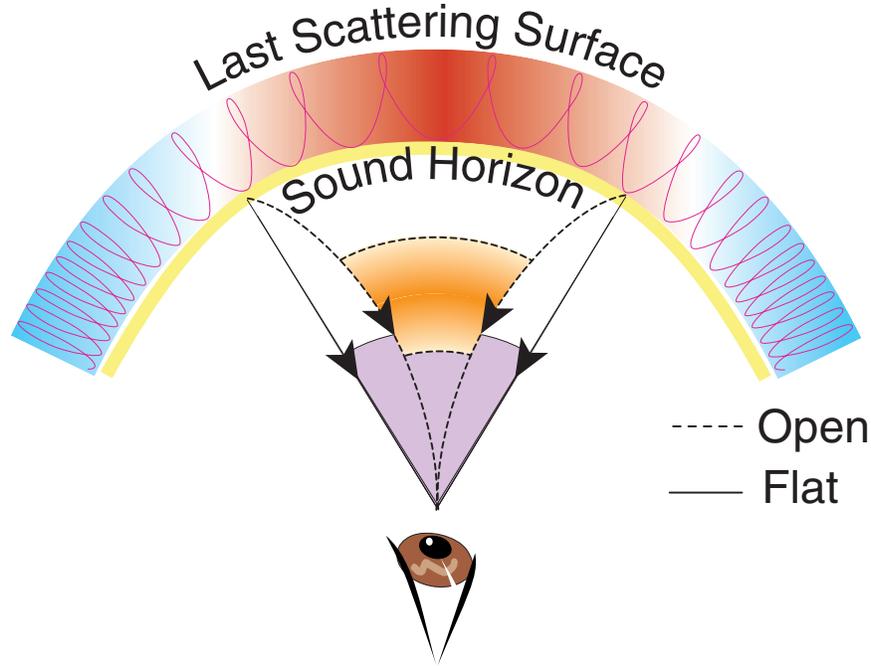,width=5in}}
\vspace{0.1in}
\caption{{\bf Gazing into the past.} 
The angle subtended on the CMB sky today by the sound horizon at
recombination depends on the various cosmological parameters. In
particular, the spatial curvature of the Universe will change the
angle under which any feature (like the sound horizon) is seen.
[adapted from a figure by Wayne Hu] }
\end{figure}

Features in the radiation pattern at this time depend on the maximum
distance a sound wave could have traveled since the Big Bang -- the
{\em sound horizon}. Cosmological models relate this distance to the
angle $\theta$ it subtends on the sky today through the
angular-diameter distance relation\cite{weinberg}. This relation
depends on the various unknown cosmological parameters, most
importantly the total energy density in the Universe. But according to
Einstein's general relativity, energy implies curvature. Hence, the
curvature of the Universe affects the angle $\theta$ subtended today
by the sound horizon at recombination (see the first figure). For a
Universe devoid of spatial curvature (a flat or Euclidean geometry)
models predict $\theta\approx 1^\circ$. Thus, if the Universe were
flat, at an angular scale of precisely $1^\circ$ we would expect to
detect some characteristic feature in the CMB, the ``fingerprint'' of
recombination.

How can this feature be detected?  One convenient way of comparing
theoretical model predictions with the result of observations is by
means of the functions $C_\ell$, the CMB angular power spectrum of the
anisotropies. The microwave sky is expanded into a set of functions
labeled by the multipole index $\ell$.  The correspondence is such
that the $\ell$th multipole samples angular scales of order
$\theta\sim 180^\circ/\ell$.  Hence, $C_\ell$ gives us the typical
strength of the temperature perturbations on that angular scale.  A
characteristic feature is given by the presence of peaks in the
$\ell(\ell+1)C_\ell$ versus $\ell$ plot. The first {\em acoustic peak} is
located at the multipole corresponding to the scale of the sound
horizon at recombination, when the plasma underwent its first
oscillation; it corresponds to a compression mode of the oscillating
plasma.

Last year, the BOOMERanG collaboration announced results from
the Antarctic long duration balloon flight mission of 1998 (B98). They
found the first peak located at $\ell\sim 200$, at the right position
for a flat Universe\cite{boom00,seife00}. Only weeks later, the results from
another balloon experiment, MAXIMA, were made available on the
internet\cite{balbi,hanany}. MAXIMA produced high-resolution maps of a 100
square-degree patch of the northern sky and went beyond B98 in
exploring multipoles from $\ell\simeq 36$ to 785, the largest range
reported to date with a single experiment.    

Recently, a joint analysis of the COBE/DMR\cite{cobe}, B98 and MAXIMA
data sets was published\cite{joint}. The COBE data provide information
at low $\ell$, necessary for normalization purposes.  After correcting
for calibration uncertainties, the B98 and MAXIMA data were quite
consistent.  The experiments used different observation strategies
and produced independent power spectra from regions of the sky roughly
$90^\circ$ apart and on opposite sides of the galactic plane.  Their
consistency gives confidence in the results (see the second figure).

\begin{figure}[htbp]
\centerline{\psfig{file=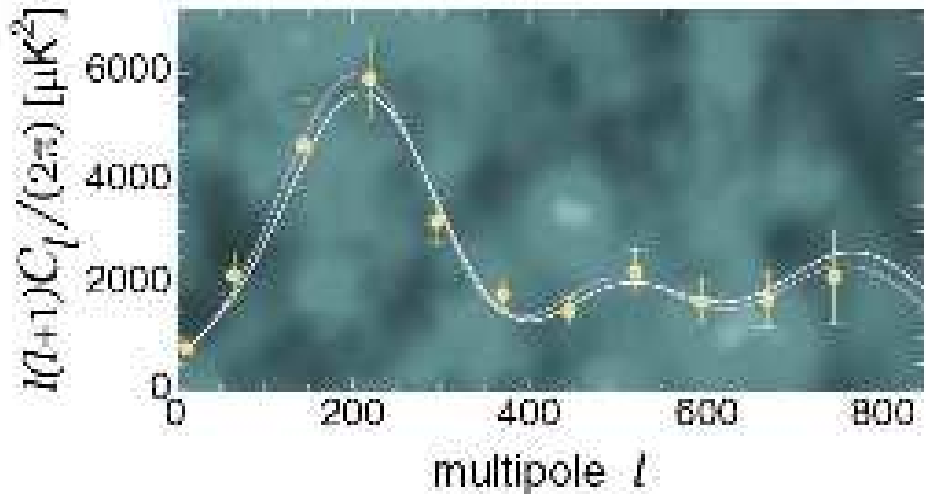,width=5in}}
\vspace{0.1in}
\caption{{\bf Distant murmurs.} The graph shows combined COBE/DMR,
BOOMERanG-98 and MAXIMA results for the CMB angular spectrum.  The
curves show the best fit model from joint parameter estimation
(pink) 
and the same but fixing a flat Universe 
(white).
These results indicate a slight excess of baryons relative to the
value determined from the relative abundance of light elements and big
bang nucleosynthesis.  The flat Universe fit becomes the best fit when
Supernova Ia data are incorporated into the analysis, implying the
existence of both non-baryonic dark matter and dark energy in the
Universe. Background: Part of the MAXIMA-1 map of CMB anisotropy. 
[adapted from Ref.[8], courtesy of Julian Borrill and Andrew Jaffe] }
\end{figure}

The presence of a localized and narrow ($\Delta\ell/\ell \sim 1$) peak
at $\ell\sim 200$ is in agreement with a flat Universe and favors an
inflationary model with initial adiabatic perturbations (where
fluctuations in each species are correlated).  In the absence of a
possible later period of reionization, which could erase partly or even
completely the acoustic peaks, the physics of recombination predicts
the existence of other peaks; the second one corresponds to a
rarefaction mode and its characteristic scale is half that of the
first peak. 
In the actual data we can see that following the first peak there is a
hint of a second one at $\ell\sim 500$, but nothing conclusive can be
said yet.

Alternative models cannot reproduce these observations.  
%sorry guys
Cosmic topological defects in their simplest versions do {\em not}
predict the existence of this oscillation pattern.  Topological
defects, such as cosmic strings and textures, which are concentrations
of primordial energy issued from early cosmological phase transitions,
can produce structure in the Universe, but cannot fit the present
data\cite{du,tu}. The complicated nonlinear evolution of the defect
network continuously perturbs the radiation background all along the
photon's journey in an incoherent fashion, leaving as its sole
characteristic signature a broad hump and virtually no acoustic
peaks\cite{Magueijo_etal96}. Recent computer simulations with a cosmic
string model\cite{levon} have failed to generate the level of CMB
variations observed by B98 and MAXIMA on scales below $1^\circ$.

The accurate locations and amplitudes of the expected secondary peaks
will allow the determination of many fundamental cosmological
parameters, such as a possible cosmological constant $\Lambda$ or
other forms of dark energy, such as
quintessence\cite{Bahcall_etal_99}.  Full analysis of the B98 and
MAXIMA data sets will provide further insights, but conclusive results
will require inclusion of CMB polarization data\cite{hedman} and full
sky coverage from the forthcoming satellite-based mission
MAP\cite{map}.  Other astrophysical input, such as supernovae and
large scale structure data, in combination with the CMB, has proven
very successful for removing degeneracies in the determination of
fundamental parameters and will be even more important in the future.

The increasing precision of today's detectors demands theoretical
modeling to be highly accurate.  The CMB contains a wealth of
information on cosmology, and future experiments will test our models
of structure formation to the limit.

\vspace{0.5in}

\begin{quote}
{\em L'int\'er\^et que j'ai \`a croire une chose} \par
{\em n'est pas une preuve de l'existence de cette chose.} \par
$~~~~~~~~~~~~~~~~~~~~~~~~$     --Voltaire, Lettres Philosophiques
\end{quote}


\begin{references}

\bibitem{Scott_etal_Science} 
D.~Scott, et al., Science 268, 829 (1995).

\bibitem{weinberg} S.~Weinberg, 
preprint arXiv.org/abs/astro-ph/0006276.

\bibitem{boom00} 
P.~de~Bernardis, et al., Nature 404, 955 (2000). 

\bibitem{seife00}
C.~Seife, Science 288, 595 (2000).

\bibitem{balbi}
A.~Balbi,  et al., Astrophys. J. 545, L1 (2000), astro-ph/0005124.  

\bibitem{hanany}
S.~Hanany, et al., Astrophys. J. 545, L5 (2000), astro-ph/0005123.

\bibitem{cobe} C.~Bennet, et al., Astrophys. J. 464, L1 (1996).

\bibitem{joint} A.~Jaffe, et al., Phys. Rev. Lett. in press (2001), astro-ph/0007333.

\bibitem{du}
R.~Durrer, A.~Gangui and M.~Sakellariadou, Phys. Rev. Lett. 76, 579 (1996).

\bibitem{tu} 
U.-L.~Pen, U.~Seljak and N.~Turok, Phys. Rev. Lett. 79, 1611 (1997); 

\bibitem{Magueijo_etal96} 
J.~Magueijo, et al., Phys. Rev. Lett. 76, 2617 (1996).

\bibitem{levon}
L.~Pogosian, in Proceedings of DPF2000, astro-ph/0009307. 

\bibitem{Bahcall_etal_99} 
N.~Bahcall, et al., Science 284, 1481 (1999).

\bibitem{hedman}
M.~Hedman, et al., Astrophys. J. submitted (2000), astro-ph/0010592.

\bibitem{map} 
http://map.gsfc.nasa.gov

\bibitem{support} 
The author is supported by CONICET and UBA (Argentina).

\end{references}
\end{document}